\documentstyle[preprint,prc,aps,floats,epsf]{revtex}
\tightenlines

\def\fpi{f_\pi}
\def\grad{{\bbox{\nabla}}}

\begin{document}
%
% Begin: Simple substitution macros used in the text
%

% uncomment \draft to have PACS numbers appear
\draft

% put preprint numbers.  
\preprint{OSU--97-232}

\title{The Skyrme Energy Functional and Naturalness}

\author{R. J. Furnstahl and James C. Hackworth}
\address{Department of Physics \\
         The Ohio State University,\ \ Columbus, OH\ \ 43210}
\date{DRAFT: \today}
\date{July, 1997}
\maketitle
\begin{abstract}
Recent studies show that successful
relativistic mean-field models of nuclei
are consistent with naive dimensional analysis
and naturalness, as expected in low-energy effective field theories
of quantum chromodynamics.
The nonrelativistic Skyrme energy functional is found to have similar
characteristics. 
\end{abstract}

\pacs{PACS number(s): 21.30.Fe, 21.60.-n, 24.85.+p}

%***************************************************************************

In a recent article \cite{FRIAR96},
evidence for quantum chromodynamic (QCD) and chiral symmetry
scales in a relativistic 
point-coupling model
of nuclei was found by examining the parameters for naturalness.
Naturalness means that coefficients of terms in the lagrangian are of
order unity after appropriate 
combinations of strong interaction scales are extracted.
Subsequent analyses of more general relativistic point-coupling 
and meson-nucleon lagrangians support this result and
give new insight into the phenomenological success of these
models \cite{FURNSTAHL97,RUSNAK97}.
One might wonder if naturalness is apparent only in relativistic models.  
In this report, we analyze a long-established {\it
nonrelativistic\/} point-coupling model: the Skyrme-force energy functional.

The Skyrme interaction \cite{SKYRME56} 
has been successfully used 
in nonrelativistic nuclear structure calculations
for many years \cite{VAUTHERIN72,NEGELE72,RING,FLOCARD,FRIEDRICH86}.
The Skyrme potential takes the form of 
zero-range (``point-coupling'') terms representing 
an expansion in the nucleon density and momentum,
and is designed for use in Hartree-Fock calculations.
It is generally interpreted as parametrizing a density-matrix
expansion of the in-medium G-matrix \cite{NEGELE72,NEGELE82}, although
in practice
the parameters are determined from direct fits to nuclear observables.

The Skyrme approach was originally proposed long before QCD and
has never been associated with QCD or chiral symmetry.
Nevertheless, it has a form {\it consistent\/} with chiral effective 
field theories 
of QCD, such as chiral perturbation theory (ChPT),
in which
the degrees of freedom are pions and nonrelativistic nucleons.
In particular,
when non-pionic degrees of freedom are integrated out, 
one expects contact terms built from powers and derivatives of the nucleon
fields, as in the Skyrme interaction.
While there are no explicit pions in the Skyrme force,
direct pion contributions largely average to zero for the bulk
properties of nuclei and the effects of pion loops 
can be approximately
absorbed into a general density functional for the energy
\cite{KOHN65,DREIZLER90}.
Thus the nucleon terms should dominate the physics of closed-shell
nuclei.

The signature of the underlying short-range physics should be the size
of the coefficients of the effective lagrangian.
However, it is not obvious that a Hartree-Fock energy functional
fit directly to finite nuclei 
should exhibit naturalness, because 
many-body correlation effects will also be absorbed into its
coefficients.
While the results from
relativistic mean-field models are encouraging, their naturalness
might rely on the large isoscalar scalar and vector
mean fields, which leads to
``Hartree dominance'' \cite{FURNSTAHL97}.

We apply
Georgi and Manohar's
naive dimensional analysis (NDA) \cite{GEORGI84b,GEORGI93}, 
which predicts the size of the coefficient of 
any term in an effective lagrangian for the strong
interaction.
This NDA has been extended to effective hadronic lagrangians for 
nuclei, both for point-coupling \cite{FRIAR96} and meson-exchange
\cite{FURNSTAHL97} models.
The basic assumption of ``naturalness'' is that once the appropriate
dimensional scales have been extracted using NDA, 
the remaining overall dimensionless
coefficients should all be of order unity.
For the strong interaction, there are two relevant scales:
the pion-decay constant $\fpi \approx 93~\mbox{MeV}$ 
and a larger scale $0.5 \alt \Lambda \alt 1~\mbox{GeV}$, 
which characterizes the mass scale of physics beyond Goldstone bosons.

The NDA rules prescribe how these scales should appear in a given 
term in the lagrangian density if it is to have a consistent loop expansion.
For a model with only nucleon fields $\psi$, 
the counting reduces
to a factor of $1/{\fpi^2\Lambda}$ for every bilinear $\psi^\dagger\psi$,
a factor of $1/\Lambda$ for every gradient, and an overall factor
of $\fpi^{2}\Lambda^{2}$.%
\footnote{In this work, we follow Ref.~\cite{FRIAR96} and do not include any
explicit counting factors.  Such factors were included in the analysis
of meson models in Ref.~\cite{FURNSTAHL97} but were not needed in
Ref.~\cite{RUSNAK97}.} 
Thus an individual term in the effective lagrangian can be written 
schematically as
\begin{equation}
     c\, \biggl[{\psi^{\dagger}\psi\over \fpi^{2}\Lambda} \biggr]^{l}
     \,  \biggl[{\grad\over\Lambda}\biggr]^{n}
     \,  \fpi^{2}\Lambda^{2} \label{eq:generic}
     \ ,
\end{equation}
with $c$ a dimensionless constant of order unity if the term is natural.
The appropriate mass for $\Lambda$ 
might be the nucleon mass $M$ or a non-Goldstone
boson mass, so  we expect $500\,\mbox{MeV} < \Lambda
< 1000\,\mbox{MeV}$.

One might try to reformulate the Skyrme approach in the form of
an effective lagrangian.
Instead, we work here
with the Skyrme energy functional,
which is most directly connected to the nuclear input.
We postulate that the size of coefficients in the
functional should be consistent with NDA.
That is, we  assume that the dominant scales of the coefficients 
are determined by the short distance physics.
A direct analysis of the Skyrme potential will be considered elsewhere.
We echo the discussion in 
Ref.~\cite{FRIAR96} and argue that
refining the Skyrme approach by adding pion loops or a more complete set
of terms will only
change values of the coefficients in the effective lagrangian
(and hence the energy functional) by
factors of order unity.
Once again, it is not at all obvious that 
many-body effects absorbed into parameters by fits to nuclei
will not disrupt the power counting;
here we test these assumptions empirically.

Therefore,
we perform our analysis on the Skyrme energy density $H({\bf r})$, which is
derived by taking the expectation value of the Skyrme hamiltonian
with respect to a Slater determinant of single-particle nucleon
wave-functions for $N=Z$ 
nuclei \cite{VAUTHERIN72,RING}.
(The energy functional itself is $\int\!d^{3}r\,H({\bf r})$.)
The result is
\begin{eqnarray}
  H({\bf r}) &=& {1\over 2M}\tau + {3\over 8} t_0 \rho^2
  + {1\over 16} t_3 \rho^3
 + {1\over 16}(3 t_1 + 5 t_2) \rho \tau  \nonumber
  \\ & & \null
  + {1\over 64} (9t_1 - 5t_2) (\grad \rho)^2  
  - {3\over 4} W_0 \rho \grad\cdot{\bf J}
  + {1\over 32}(t_1-t_2) {\bf J}^2  \ ,  \label{eq:H}
\end{eqnarray}
where $\rho({\bf r})$ is the nucleon density, 
$\tau({\bf r})$ is the kinetic energy
density, and ${\bf J}({\bf r})$ is the so-called spin-orbit 
density \cite{RING}.
Some other variations of the Skyrme interaction lead to fractional powers
of $\rho$ in $H({\bf r})$ \cite{FRIEDRICH86} 
and will not be considered here.   
The coefficients in $H({\bf r})$ are 
determined by fits to
nuclear observables; it is the analog of the energy functionals
in the 
relativistic mean-field analyses.

\def\mc#1{\multicolumn{1}{c}{$\quad #1$}}
\def\zz{\phantom{1}}
\def\zm{\phantom{\mbox{$-$}}}

\begin{table}[t]
\caption{Parameter sets for some standard Skyrme interactions.
}
%\smallskip
\vspace{.1in}
\begin{tabular}[t]{ccccccc}
   & $t_0$  & $t_1$  & $t_2$ & $t_3$ & $W_0$ & $x_0$ \\
 Force   & (MeV-fm$^3$) & (MeV-fm$^5$) & (MeV-fm$^5$) & (MeV-fm$^6$) &
          (MeV-fm$^5$) &   \\ 
   \hline
  Skyrme 1  & $-1057.3$  & 235.9 & $-100.0$  & 14463.5  & 120 
           & \zz 0.56 \\
  Skyrme 2  & $-1169.9$  & 585.6 & $\zz -27.1$  & \zz 9331.1  & 105 
           & \zz 0.34 \\
  Skyrme 3  & $-1128.8$  & 395.0 & $\zz -95.0$  & 14000.0  & 120 
           & \zz 0.45 \\
  Skyrme 4  & $-1205.6$  & 765.0 & $\zm\zz 35.0$  & \zz 5000.0  & 150 
           & \zz 0.05 \\
  Skyrme 5  & $-1248.3$  & 970.6 & $\zm 107.2$  & \zz\zz\zz\zz 0.0  & 150 
           & \zz 0.17 \\
  Skyrme 6  & $-1101.8$  & 271.7 & $-138.3$  & 17000.0  & 115 
           & \zz 0.58 \\
\end{tabular}
\label{tab:one}
\end{table}

\begin{table}[tb]
\caption{Dimensionless coefficients obtained
 for some conventional Skyrme interactions
 by applying Eq.~\ref{eq:scaled}.
}
%\smallskip
\vspace{.1in}
\begin{tabular}[t]{cccccccc}
 Force  & $c_1$  & $c_2$  & $c_3$ & $c_4$ & $c_5$ & $c_6$ & $c_7$ \\
   \hline
  Skyrme 1 & 0.5 & $-0.45$ & 1.09 & 0.33 & 1.05 & $-2.31$ & 0.27  \\
  Skyrme 2 & 0.5 & $-0.50$ & 0.70 & 2.60 & 2.17 & $-2.02$ & 0.49    \\
  Skyrme 3 & 0.5 & $-0.48$ & 1.05 & 1.14 & 1.62 & $-2.31$ & 0.39    \\
  Skyrme 4 & 0.5 & $-0.51$ & 0.38 & 3.97 & 2.70 & $-2.89$ & 0.59    \\
  Skyrme 5 & 0.5 & $-0.53$ & 0.00 & 5.53 & 3.29 & $-2.89$ & 0.69    \\
  Skyrme 6 & 0.5 & $-0.47$ & 1.28 & 0.20 & 1.26 & $-2.21$ & 0.33    \\
\end{tabular}
\label{tab:two}
\end{table}

For the purpose of applying Eq.~(\ref{eq:generic}), 
we make the correspondences (neglecting
irrelevant signs and spin matrices):
\begin{eqnarray}
   \rho &\longleftrightarrow& \psi^\dagger\psi \ , \label{eq:rho} \\
   \tau &\longleftrightarrow& \grad\psi^\dagger\cdot\grad\psi \ , 
                 \label{eq:tau}  \\
   {\bf J} &\longleftrightarrow& \psi^\dagger\grad\psi \ .  \label{eq:J}
\end{eqnarray}
Applying the scaling rules from (\ref{eq:generic})
term by term to Eq.~(\ref{eq:H}), 
we can rewrite $H({\bf r})$ in terms of
dimensionless coefficients $c_i$, which should be of order one if
natural:
\begin{eqnarray}
  H({\bf r}) &=&
   c_1{\tau\over \Lambda} + c_2 {\rho^2\over \fpi^2}
  + c_3 {\rho^3\over \fpi^4\Lambda} 
 + c_4 {\rho \tau\over \fpi^2\Lambda^2}
  + c_5 {(\grad \rho)^2 \over \fpi^2\Lambda^2}  
  + c_6 {\rho \grad\cdot{\bf J} \over \fpi^2\Lambda^2}
  + c_7 {{\bf J}^2 \over \fpi^2\Lambda^2}  \nonumber \\
          &=&
  {\tau\over \Lambda}\Bigl( c_1 +  c_4 {\rho\over \fpi^2\Lambda}\Bigr)
   + {\rho^2\over \fpi^2}\Bigl( c_2 
  + c_3 {\rho\over \fpi^2\Lambda}\Bigr) 
  + c_5 {(\grad \rho)^2 \over \fpi^2\Lambda^2}  
  + c_6 {\rho \grad\cdot{\bf J} \over \fpi^2\Lambda^2}
  + c_7 {{\bf J}^2 \over \fpi^2\Lambda^2} \ .
     \label{eq:scaled}
\end{eqnarray}
The second line manifests the expansion and truncation of
$H({\bf r})$ in powers and derivatives of the nucleon
fields, with expansion parameter $\rho/\fpi^{2}\Lambda$.

There are many sets of parameters
for the Skyrme force determined by different fits to nuclear 
observables such as experimental binding energies and radii of nuclei.
Here we consider Skyrme~1 through Skyrme~6 \cite{FRIEDRICH86},
which are parameters for models with energy densities of the form
of $H({\bf r})$ in Eq.~(\ref{eq:H}). 
The coefficients $t_{0}$ through $t_{3}$ and $W_{0}$ are dimensional, 
but as usually presented there is little clue to the 
relevant scales that determine their size.
In Table~\ref{tab:one}, we list the coefficients for models 1 through 6,
in the conventional units; the coefficients are certainly not
natural as given!

Coefficients for Skyrme~1 through 6 scaled according to
Eq.~(\ref{eq:scaled}) are given in Table~\ref{tab:two}, 
where we have used $\Lambda = M$.
While this is likely an upper limit to $\Lambda$ in some cases, 
the coefficients for
$500\,\mbox{MeV} < \Lambda < 1000\,\mbox{MeV}$ are not qualitatively 
different ($\Lambda = 1000\,\mbox{MeV}$ was used in 
Ref.~\cite{FRIAR96}).
However, the range in $\Lambda$ 
increases the uncertainty when estimating the size
of omitted higher-order contributions (see below).
 
An obvious example of a natural coefficient is $c_1 = 1/2$, which
follows since the scale of nucleon kinetic energy is the nucleon mass.
However, the scaling of the other coefficients is non-trivial;
if $M$ alone were extracted to define dimensionless coefficients 
they would be badly unnatural.
For example, $c_{3}$ would be over $10^{5}$ for most of the forces.
The coefficients are also unnatural if one expressed $H({\bf r})$ as an
expansion in $\rho({\bf r})/\rho_{0}$, where $\rho_{0}$ is the
saturation density of nuclear matter.

In contrast, the NDA scaling of Eq.~(\ref{eq:scaled}) implies natural
coefficients in essentially all cases.
The ``worst case'' is $c_{4}$ in Skyrme~5, but
we also note that this interaction is particularly unnatural by
construction, since $c_{3}$ is taken to be zero.

\begin{figure}[p]
 \setlength{\epsfxsize}{4.0in}
  \centerline{\epsffile{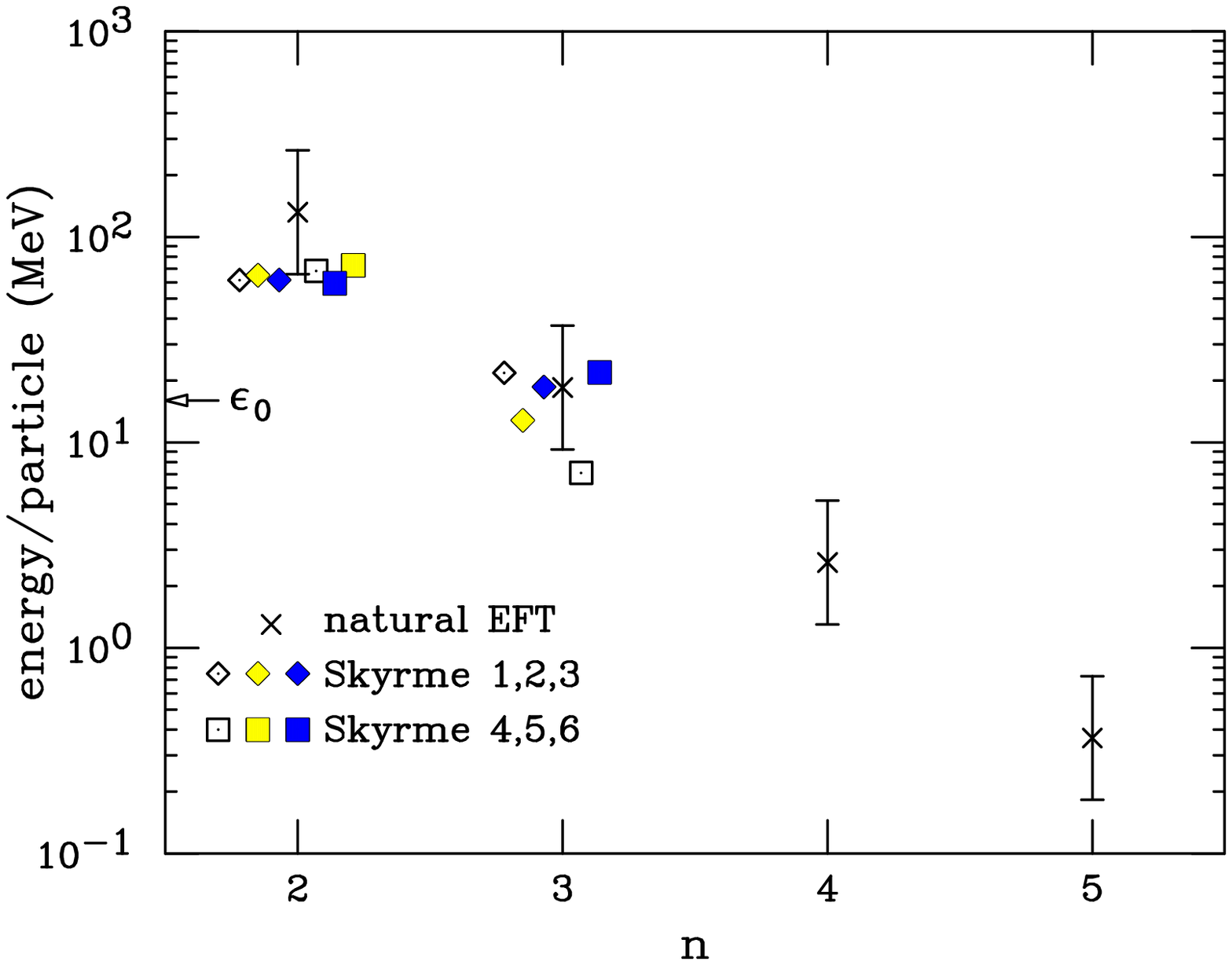}}
\vspace{.1in} \caption{\small Contributions to the energy
 per particle in nuclear matter from terms
 of the form $\rho^{n}$, evaluated at saturation density $\rho_{0}$ 
 for a variety of  Skyrme interactions.  
The crosses are estimates based on Eq.~(\ref{eq:generic})
with $\Lambda = 939\,\mbox{MeV}$.
The arrow indicates the total binding energy $\epsilon_0 =16.1~\mbox{MeV}$.}
 \label{fig:one}
\end{figure}

Figure~\ref{fig:one} shows the contributions
to the nuclear matter energy
per particle of the form $\rho^{n}$,
evaluated at saturation density $\rho_{0}$.
The crosses are estimates based on the assumption of natural
coefficients given by Eq.~(\ref{eq:generic}) with 
$\psi^{\dagger}\psi \rightarrow \rho_{0}$, 
and the error bars show a range from
1/2 to 2 in the coefficients  
($\Lambda=M$ is used in Fig.~\ref{fig:one}).
The Skyrme contributions are consistent with naturalness, although
a more systematic study of Skyrme-type energy functionals including
higher powers of $\rho$ would be needed to be conclusive.

It is evident that naturalness implies a convergent density expansion
for mean-field contributions to nuclear matter, with expansion
parameter $\rho_{0}/\fpi^{2}\Lambda$ between 1/4 and 
1/7 \cite{FRIAR96b}.
One can also anticipate good convergence for terms with gradients
of the fields, since the nucleons are nonrelativistic, and gradients
of the densities, since the relevant scale for derivatives 
in finite nuclei should
be roughly the nuclear surface thickness $\sigma$, and so the predicted
dimensionless expansion parameter is $1/\Lambda\sigma \leq 1/5$.

Nevertheless, the initial energy scale is large compared to the nuclear
binding energy so that the $n=3$ term is still important.
(The size of the $n=2$ contribution is discussed below.)
The largeness of this ``three-body'' term 
is conventionally cited \cite{RING}
as implying a strong density dependence to the microscopic effective
interaction.%
\footnote{There are many sources of such terms in a 
nonrelativistic effective lagrangian, including 
relativistic effects \cite{FOREST95,SEROT97}.} 
While $\rho^{5}$ terms are unlikely to be relevant,
the omitted $\rho^{4}$ contribution is estimated to be 
uncomfortably large (and would be larger with a smaller value of 
$\Lambda$) at
nuclear saturation density.

\begin{figure}[p]
 \setlength{\epsfxsize}{4.0in}
  \centerline{\epsffile{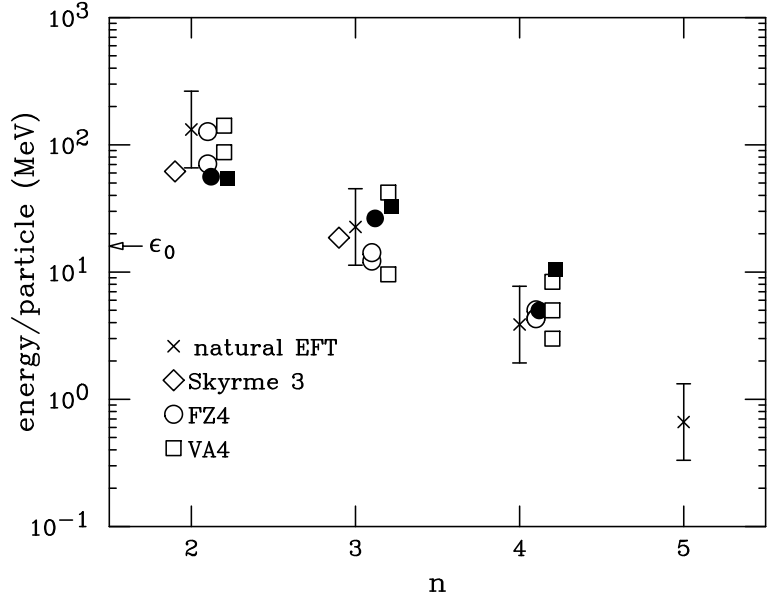}}
\vspace{.1in} 
\caption{\small Contributions to the energy
 per particle in nuclear matter at saturation density from terms
 of the form $\rho^{n}$ for the Skyrme~3 model
 and $\rho_{{\rm s}}^{i}\rho_{{\rm v}}^{j}$ with $i+j=n$ for two relativistic
 point-coupling models from Ref.~\protect\cite{RUSNAK97} (see text).  
The crosses are estimates based on Eq.~(\ref{eq:generic})
with $\Lambda = 770\,\mbox{MeV}$.
The arrow indicates the total binding energy $\epsilon_0 =16.1~\mbox{MeV}$.}
 \label{fig:two}
\end{figure}

In Figure~\ref{fig:two} we compare the typical Skyrme result (Skyrme 3
is used) to results from general relativistic point-coupling models
fit to nuclear observables \cite{RUSNAK97}.
Contributions from individual terms to two relativistic models 
(labeled FZ4 and VA4) are shown as unfilled circles and squares while
the net contributions are shown as filled symbols.%
\footnote{Absolute values are plotted in the figure.}
The multiple contributions for each $n$ in the relativistic 
models are of the form $\rho_{{\rm s}}^{i}\rho_{{\rm v}}^{j}$ with $i+j=n$,
where $\rho_{{\rm s}}$ is the scalar density and $\rho_{{\rm v}}$ 
is the vector
(baryon) density. 
The naturalness of the relativistic models implies an expansion that
can be truncated at $n=4$ with an error of order 1~MeV, which is
easily absorbed by slight adjustments of the other parameters.

The strong cancellation between the $\rho_{{\rm s}}^{2}$ and
$\rho_{{\rm v}}^{2}$ terms is characteristic of relativistic point-coupling
models \cite{RUSNAK97}.  
A nonrelativistic reduction of the point-coupling model would
incorporate this cancellation and therefore can be anticipated in the
Skyrme energy.%
\footnote{The correspondences between the Skyrme energy functional and 
relativistic mean-field models has been discussed by Reinhard 
and collaborators \cite{REINHARDT}.}
Indeed, the Skyrme $n=2$ energy is consistent with the {\it net\/} $n=2$
contribution from the relativistic models, which is just marginally
natural because of the cancellations.
For higher-order terms, however, the net contribution 
is comparable to individual
contributions, so one cannot rely on further cancellations to improve the
convergence of the nonrelativistic expansion.
Thus if the NDA estimates are used to anticipate contributions in
a complete nonrelativistic point-coupling model,
it would appear that $n=4$ contributions are still significant
at nuclear saturation density.

Note that this conclusion does not contradict the conventional wisdom
from few-body calculations that four-body contributions are quite 
small \cite{PUDLINER,FRIAR96b},
because the effective densities involved are significantly lower.
Furthermore, while
$n=4$ terms in relativistic 
mean-field {\it meson\/} models are important for achieving good fits 
to bulk nuclear observables \cite{FURNSTAHL97},
very good fits can be obtained in point-coupling models with a truncation
at $n=3$.
(In both cases the best fits require $n=4$.)
The $n \le 3$ coefficients are able to adjust to absorb to a large
degree the higher-order contributions.
Thus it is not surprising that the Skyrme energy functional in its
usual form is successful in reproducing nuclear observables.

Because the Skyrme
energy functional includes only a limited set of terms,
our results here are not by themselves definitive. 
But in the context
of the other more complete investigations of relativistic models 
they are quite encouraging.
In future work we will make
a more extensive evaluation of Skyrme-like
forces 
using the same approach applied to relativistic models 
\cite{FURNSTAHL97,RUSNAK97}.
This means considering all possible nonredundant terms (consistent
with symmetries) in the energy functional, organized according
to NDA. 
The goal is to constrain the parameters using a wide range of
observables, instead
of minimizing the number of parameters to improve predictability.
The connection between naturalness in an effective Skyrme-like
lagrangian and
naturalness in the implied nonrelativistic Hartree-Fock energy functional
(and beyond) will also be explored.

In summary, we have examined the 
nonrelativistic Skyrme energy functional in the context of low-energy 
effective field theories of QCD.
As was found for relativistic point-coupling and mesonic models,
Skyrme
parameters are natural after applying naive dimensional analysis.
This implies that QCD scales are relevant in analyzing the physics
of nuclei,
despite the complicated many-body physics and subtle dynamics of
nuclear saturation that are absorbed into the parameters of
the energy functional.
The NDA provides a new organizational principle for Skyrme-like
models at the mean-field level that suggests that current
models are truncated prematurely.
These results encourage the further application of effective field theory
methods to finite density nuclear systems.

%***************************************************************************

\acknowledgments

We thank S.~Brand, J.~Rusnak, and B.~Serot for useful discussions.
This work was supported in part by the 
National Science Foundation
under Grants No.\ PHY--9511923 and PHY--9258270.

%%%%%%%%%%%%%% FIGURES %%%%%%%%%%%%%%%%%%%%%%%%%%%%%%%%

\end{document}